\documentclass{emulateapj}
\usepackage{natbib,graphicx}

\def\bnabla{\mbox{\boldmath $\nabla$}}

\def\be{\begin{equation}}
\def\ee{\end{equation}}

\def\ergs{\, \rm ergs}
\def\s{\, \rm s}
\def\K{\, \rm K}

\def\s{\, \rm s}

\def\pomega{\varpi}
\def\deg{^\circ}

\usepackage{ulem}  
\usepackage{color}

\begin{document} 
	
\title{\mbox{Secular Chaos and the Production of Hot Jupiters}}
\author{Yanqin Wu\altaffilmark{1}, Yoram Lithwick\altaffilmark{2,3}}
\altaffiltext{1}{Dept. of Astronomy \& Astrophysics, University of
  Toronto, Toronto, ON, Canada} \altaffiltext{2}{Canadian Institute of
  Theoretical Astrophysics, Toronto, ON, Canada}
\altaffiltext{2}{Department of Physics \& Astronomy, Northwestern
  University, Evanston, IL, USA}

\begin{abstract}
  In a planetary system with two or more well-spaced, eccentric,
  inclined planets, secular interactions among these planets may lead
  to chaos.   
The innermost planet may gradually become increasingly
eccentric and/or inclined, 
 as a result of the
 secular degrees of freedom drifting towards  equipartition
 of AMD
  (angular momentum deficit).
  This ``secular chaos'' is  known to be responsible for the eventual
  destabilization of Mercury in our own Solar System. Here we focus on
  systems with three giant planets.
We characterize the secular chaos and demonstrate the criterion
for it to occur, but leave a detailed understanding of secular
chaos to a companion paper \citep{yoram}. 

After an extended period of eccentricity diffusion, the inner planet's
pericentre can approach the star to within a few stellar radii. Strong
tidal interactions and ensuing tidal dissipation extracts orbital
energy from the planet and pulls it inward, creating a hot Jupiter.
In contrast to other proposed channels for the production of hot
Jupiters, such a scenario (which we term ``secular migration'')
provides an explanation for a range of observations: the pile-up of
hot Jupiters at 3-day orbital periods, the fact that hot Jupiters are
in general less massive than other RV planets, that they may have
misaligned inclinations with respect to stellar spin, and that they
have few easily detectable companions (but may have giant companions
in distant orbits). Secular migration can also explain close-in
planets as low in mass as Neptune;
{ and an aborted secular migration can explain the ``warm Jupiters''
  at intermediate distances.  In addition,} the frequency of hot
Jupiters formed via secular migration increases with stellar age.  We
further suggest that secular chaos may be responsible for the observed
eccentricities of giant planets at larger distances, and that these
planets could exhibit significant spin-orbit misalignment.
\end{abstract}

\setcounter{equation}{0}

\section{Introduction}
\label{sec:intro}

While around $10\%$ of sun-like stars surveyed harbor Jovian-mass
planets, only $\sim 1\%$ are orbited by so-called hot-Jupiters with
periods short-ward of $\sim 10 $ days \citep[see reviews
by][]{Marcy,Udry}. There appears to be a pile-up of hot Jupiters
around $3$ day orbital periods. This excess is genuine and has been
confirmed by both radial velocity and transit surveys
\citep{Gaudi,Butler,Cumming,Fressin}.  Outward of hot Jupiters, there
appears a deficit of gas giants with periods of $10$ to $100$ days
\citep[the ``period valley;''][]{UdryMayorSantos,Wittenmyer}.

According to current theories of planet formation, hot Jupiters could
not have formed in situ,  given the large stellar tidal field, high
  gas temperature, and low disk mass to be found so close to the star.
 Instead, hot Jupiters most likely formed beyond a few AU and then
migrated inward.  Candidate migration scenarios that have been
proposed  include protoplanetary disks, Kozai migration by binary
or planetary companions,  and scattering with other planets in the
system. While each of these mechanisms may have contributed to the hot
Jupiter population to some degree, the question remains as to which is
the dominant  one. The dominant mechanism has to explain a variety
of observed correlations. In \S \ref{sec:critical}, we  review
some of these correlations and provide a critical assessment of the
above three mechanisms.

In this work, we propose a fourth channel for producing hot Jupiters,
namely planet migration by secular chaos. Secular chaos may arise in
planetary systems that are well-spaced and are dominated by long-range
secular interactions.   A system of two non-coplanar planets can be
  chaotic, but only if their eccentricities and inclinations are of
  order unity  \citep{Libert,Migas09,Smadar}.  So in this
contribution we focus on systems with three planets. The criterion for
secular chaos  is less stringent, and the character of secular
chaos is more diffusive, differing from that of the two planet case.
This diffusive type of secular chaos promotes energy equipartition
between different secular degrees of freedom.  The physics behind
secular chaos is analyzed in detail in a companion paper
\citep{yoram}, where we show that Mercury, the innermost planet in our
Solar system, experiences a similar type of secular chaos. Mercury may
consequently be removed from the Solar system
\citep{Laskar08,Batygin,LaskarGastineau}.

Secular chaos tends to removes angular momentum in the inner most
planet gradually, causing its pericenter to approach the star.  Tidal
dissipation may then remove orbital energy from this planet,
  turning it into a hot Jupiter. Hot Saturns or hot Neptunes may also
be produced similarly. Such a migration mechanism, which we term
``secular migration,'' can reproduce a range of observations. It also
predicts that in systems with hot planets, there are other giant
planets roaming at larger distances.

\section{Hot Jupiters: observations and theories}
\label{sec:critical}

\subsection{Observations}

There is a sharp inner cut-off to the 3-day pile-up of hot
Jupiters. They appear to avoid the region inward of {\it twice} the
Roche radius \citep{FordRasio}, where the Roche radius is the distance
within which a planet would be tidally shredded.  New data spanning
two orders of magnitude in planetary masses (and including planet
radius measurements) have strengthened this claim. There are only 5
known exceptions lying inward of twice the Roche radius, and the rest
mostly lie between twice and four times the Roche radius.

Hot Jupiters appear to be less massive than more distant planets
\citep{Patzold,Zucker}.  For planets discovered with the radial
velocity method, close-in planets have projected masses ($M\sin i$)
less than twice Jupiter's mass, excluding planets in multiple star
systems.  But numerous further out planets have $M\sin i>2M_J$
\citep[][Fig. 5]{Udry}.

Many hot Jupiters have orbits that are misaligned { with} the spin
of their host stars. The angle between the orbit normal of a
transiting planet and the spin axis of its host star (the stellar
obliquity) can be probed with the Rossiter-McLaughlin (R-M) effect
\citep[e.g.][]{Winn05}.
Planets were presumably born in a disk aligned with the stellar
spin. Therefore measurements of the present stellar obliquity provide
a stringent constraint on the migration scenario.  Analysis of the
first $11$ systems with R-M measurements found that the majority were
consistent with perfect alignment, while a small minority were highly
misaligned \citep{FabryckyWinn}.  However later analysis that included
more systems ($26$ in total) found that most are misaligned, and many
of these are even in retrograde orbits \citep{Triaud}.  The reason for
this discrepancy is currently unclear.

Hot Jupiters also tend to be alone, { at least out to} a few AU.  From
radial velocity surveys, $\sim 30\%$ of planets are in multiple planet
systems \citep[including ones with radial velocity trends,][]{Butler},
while only 5 hot Jupiters are \citep[HD 187123b, HIP 14810b, ups Andb,
HAT-P-13b and HD 217107b;][]{Wright,Hebrard}; {i.e. fewer than $10\%$
  of hot Jupiters are known to have companions within a couple AU.}
This relative deficit also shows up in the transit sample, where most
attempts at detecting transit timing variations caused by close
companions \citep{HolmanMurray,Agol} have been unsuccessful
\citep[e.g.][]{Rabus,Csizmadia,Hrudkov}, except for, perhaps,
\citet{Macie0,Macie,Fukui}.

\subsection{Migration Theories}

The successful migration scenario has to explain these and other
observed correlations. There are two categories of migration
scenarios.  One is to migrate the planet within a gaseous disk (disk
migration).  The other is to generate a high eccentricity in the
planet, bringing it sufficiently close to the star that tidal
dissipation circularizes and shrinks its orbit into that of a hot
Jupiter.  The latter category includes Kozai migration, planet-planet
scattering, as well as the secular chaos that we propose in this work.


We first examine disk migration, a theory that pre-dated the discovery
of hot Jupiters \citep{LinPapaloizou}. It asserts that the viscous
protoplanetary disk carries the planet inward
\citep{chambers09,rice08}. The presence of mean-motion resonance pairs
among observed planets seems to support this scenario. However, to
produce the observed pile-up of hot Jupiters at around 3 day orbital
periods, the inward migration has to be halted. As discussed by
\cite{Lin}, this could be achieved if the disks are truncated by the
stellar magnetosphere at a radius that corresponds to twice the planet
orbital period. Disks are likely truncated at the co-rotation radius,
where the orbital period of the disk material equals the spin period
of the star. The observed distribution of rotation periods of pre-main
sequence stars appear to be bimodal \citep{Herbst} with location of
the long period peaks varying from 4 to 8 days for different clusters.
This could be consistent with the period distribution of hot Jupiters.
It is also plausible that hot Jupiters thus migrated would tend to be
lower in mass, although that has yet to be shown.  A difficulty with
this scenario is that it should produce planets whose orbit normals
are aligned with the stellar spin axis \citep[for an opposite view,
however, see][]{LaiLin}.  In addition, it remains to be argued why hot
Jupiters rarely have close companions (within a few AU) {if they} are
migrated inward by powerful disks.  And there is no natural
explanation in this scenario for the avoidance of twice the Roche
radius.


We now turn our attention to planet-planet scattering. First proposed
by \citet{RasioFord}, it asserts that close encounters between planets
can induce extreme eccentricity in one of them
\citep{Ford01,PapaloizouTerquem,Ford08}, which may then be tidally
circularized to form a hot Jupiter. Such a theory reproduces the
observed eccentricity distribution of (non-hot) extra-solar planets
that have $e\gtrsim 0.2$ \citep{Chatterjee,JuricTremaine}, as well as
the observed close-packed nature of pairs \citep{Raymond}. It may also
account for the high inclinations of hot Jupiters.  However, it is
unclear if the initial condition of a compact and highly unstable
planetary system as required by this theory is applicable to planets
emerging from a gaseous disk \citep{Matsumura}. Also,
\cite{Chatterjee} find that the inner planets tend to be the most
massive ones, contrary to the observed correlations. Furthermore,
since scatterings are sudden, it is difficult for tides, a slow
process, to halt scattered planets \citep{Nagasawa}. This theory also
predicts readily detectable outer planets that are responsible for
scattering and producing the observed hot Jupiters. They are, however,
not observed.


Lastly, we comment on Kozai migration. First proposed by
\citet{WuMurray}, it asserts that a highly inclined companion star can
induce Kozai oscillations \citep{Kozai,Eggleton} in the planet,
gradually exciting the planet to a high enough eccentricity that it
approaches the central star, whereupon tidal dissipation circularizes
it into a hot Jupiter.  While it succeeds in { producing hot Jupiters
  that are highly inclined with respect to stellar spin,} including
ones that are retrograde in projection \citep[][]{Triaud}, and is
likely responsible in a number of specific cases \citep[such as HD
80606b][]{Naef,Laughlin,Pont,Winn,Hebrard10}, it does not
preferentially yield low-mass hot Jupiters,\footnote{Kozai migration
  can readily migrate massive planets inward, and perhaps can account
  for the presence of massive hot Jupiters found in binary systems
  \citep{Zucker}.}  and its effectiveness may be hampered by the
presence of other planets in the system \citep{WuMurray}.
Furthermore, population studies establish that only $\sim 10\%$ of hot
Jupiters can be explained by Kozai migration due to binary companions
\citep{WuMike,fabtremaine}.

Mechanisms that rely on eccentricity excitation, such as Kozai
migration or planet-planet scattering, naturally produce hot Jupiters
that tend to avoid the region inside of twice the Roche radius
\citep{FordRasio}.  However, only Kozai migration can naturally
explain the 3-day pile-up, as the eccentricity rise in this case is
gradual and planets are accumulated at the right location.  The
secular chaos mechanism described in this paper also leads to gradual
eccentricity excitation and can therefore inherit much of the success
of the Kozai mechanism.

The noteworthy simulations of \citet{Nagasawa} combine planet
scattering with Kozai oscillations.  Starting from very compact
systems of three equal-mass planets, their planets frequently scatter
one another onto highly inclined orbits, which in some cases triggers
Kozai oscillations.  Their particular set-up yields hot Jupiters $\sim
30\%$ of time, with orbital inclinations that are roughly isotropic.
The production of hot Jupiters by inter-planet Kozai oscillations has
also been studied in \cite{Smadar}.  Such a mechanism appears
promising if eccentricities and inclinations can reach order-unity
values.

\subsection{Secular Interactions}

Secular interactions are a simplified version of interplanetary
interactions, where one can account for the forces between two planets
by calculating the torque between two mass wires. The latter are made
by spreading the mass of a planet along its orbit, weighted by the
amount of time it spends at that segment. This describes the dynamics
adequately as long as the planets have no close-encounters and do not
lie near mean-motion resonances. Secular interactions allow planets to
exchange angular momentum but not energy. So planets' semi-major axes
are unchanged.  The long-term evolution of the inner Solar system, for
instance, is primarily secular in nature \citep{Laskar89}.

Under certain circumstances, secular interactions can gradually raise
the eccentricity of the inner planet to near unity, even when the
initial eccentricity are as low as that expected of planets emerging
out of dissipative gaseous protoplanetary disks.  This is the work of
chaos.

When planet eccentricities and inclinations are small, secular
dynamics is fully described by a linear summation of secular
eigenmodes that are independently oscillating, abiding by the
so-called Laplace-Lagrange theory \citep[see, e.g.][]{MD00}. The
multi-periodic variations in the eccentricity of the Earth are largely
caused by the interference between the eigenmodes.  These have been
claimed to drive climate changes \citep{Milankovitch}.

When eccentricities and inclinations rise, linear eigenmodes no longer
describe the dynamics adequately. 
Nonlinear effects can occur.  One example is the appearance of
nonlinear secular resonances, including new fixed points and
separatrices that are not present in the linear system
\citep{Malhotra,Mich06,Migas09}.  A second is the chaotic motion
associated with the overlap of neighboring resonances
\citep{sid90,Mich06,yoram}.

For a system of two planets, strong nonlinearity and chaos requires
eccentricities and/or inclinations of order unity.
If the two planets are coplanar, 
 energy and angular momentum
  conservation constrain the motion to be regular and quasi-periodic.
But if  the planets are sufficiently inclined,
 the Kozai resonance can 
be triggered, leading to instability and/or chaos.
This requires mutual inclinations $\gtrsim 40\deg$
if the initial eccentricities are very large, 
and $\gg 40\deg$ for more modest eccentricities
\citep{Mich06,Smadar}.
   For instance, for the inner two planets listed in Table
\ref{tab:initial} to interact to produce $e_1 > 0.98$, a mutual
inclination of $ > 85 \deg$ is necessary.

For a system of three or more planets, however, the threshold for
chaos is much reduced.  Moreover, the nature of the secular chaos is
different.  In the two planet case, chaotic systems are { quickly}
unstable on the secular timescale, $\sim 10$ Myr for typical
parameters \citep{Mich06,Smadar}.  By contrast, in the multi-planet
case a multitude of resonances can overlap.  This leads to a gentler
type of chaos, with the orbital elements gradually diffusing over many
secular times, while the values of the eccentricities and inclinations
remain modest.  An example of the latter type of behavior is the inner
Solar system, where chaos is prevalent even at
eccentricity/inclination levels of a few percent, and the timescale of
the evolution is $\gg$ Gyr \citep{Laskar89,yoram}.

In the following, we investigate a planetary system with three mildly
eccentric { and} inclined planets to demonstrate the appearance of
this new type of secular chaos --- new, that is, in the context of
extra-solar planetary systems.

\section{Secular Chaos: A Worked Example}
\label{sec:example}

\subsection{Numerical Example}
\label{subsec:numerics}

\begin{table}
\begin{center}
\caption{Initial conditions for the example system}
\begin{tabular}{lcccccc}
\hline
pl. & mass ($M_J$) & a(AU) & e & inc (deg) & $\omega$ (deg) & $\Omega$ (deg) \\
\hline \\  
$1$ & $0.5$ & $1$ & $0.066$ & $4.5$ & $\pi$ & $0$ \\
$2$ & $1.0$ & $6$ & $0.188$ & $19.9$ & $0.38\pi$ & $\pi$ \\
$3$ & $1.5$ & $16$ &  $0.334$ & $7.9$ & $\pi$ & $0$ \\
\hline
\label{tab:initial}
\end{tabular}
\end{center}
\end{table}

{\bf Initial conditions:} Parameters for the example system we
investigate are listed in Table \ref{tab:initial}. The inclinations
are measured relative to the system's invariable plane.

For the semi-major axes, we space the  planets sufficiently
  far apart and away from any major mean-motion resonances (orbital
period ratios are: $1:14.7:64$). The choice for the masses  is
somewhat arbitrary, except for our choice that the innermost
  planet be the least massive, which facilitates its excitation.

The somewhat odd-looking choices for the other orbital elements 
  place most of the secular energy (i.e., angular momentum
deficit) in the outer planets, or more specifically in the secular
eigenmodes associated with  those planets (more detail in \S
\ref{sec:analy}). We find that the occurrence of secular chaos is not
particularly sensitive to our choices  for these values, as long
as the system has sufficient amount of angular momentum deficit (\S
\ref{sec:analy}).

{\bf Numerics:} This simulation was performed using the SWIFT
symplectic integrator \citep{levisonduncan}, supplemented with
routines that model tidal dissipation in the planet { and in the
  star}, precessions due to general relativistic effects, and
precessions due to rotational and tidal bulges on both the inner
planet and the star.  Details are presented in the Appendix. These
effects are essential for determining the final positions of the hot
Jupiters.

Numerical precision at extremely high eccentricities is of concern.
So we adopt a time-step that is $1/100$ of the inner planet's orbital
period (1 yr) for most of the integration, but switch it to a value
$100$ times shorter whenever the periastron of the inner planet
reaches inward of $0.1$ AU from the star. We find that such a change
of time-step, even though it breaks the time-symmetry of the
symplectic integrator, is required to maintain satisfactory energy and
angular momentum conservation.
The fractional energy error, integrated over an episode of extremely
high eccentricities (which typically lasts $\sim 10^4$ yrs), remains
below $10^{-4}$ as long as $e_1 < 0.98$, sufficiently small for our
problem at hand.

 The angular momentum error is much smaller.
\begin{figure*}
\begin{center}
\includegraphics[width=0.45\textwidth,angle=-90,
trim=300 100 50 120,clip=true]{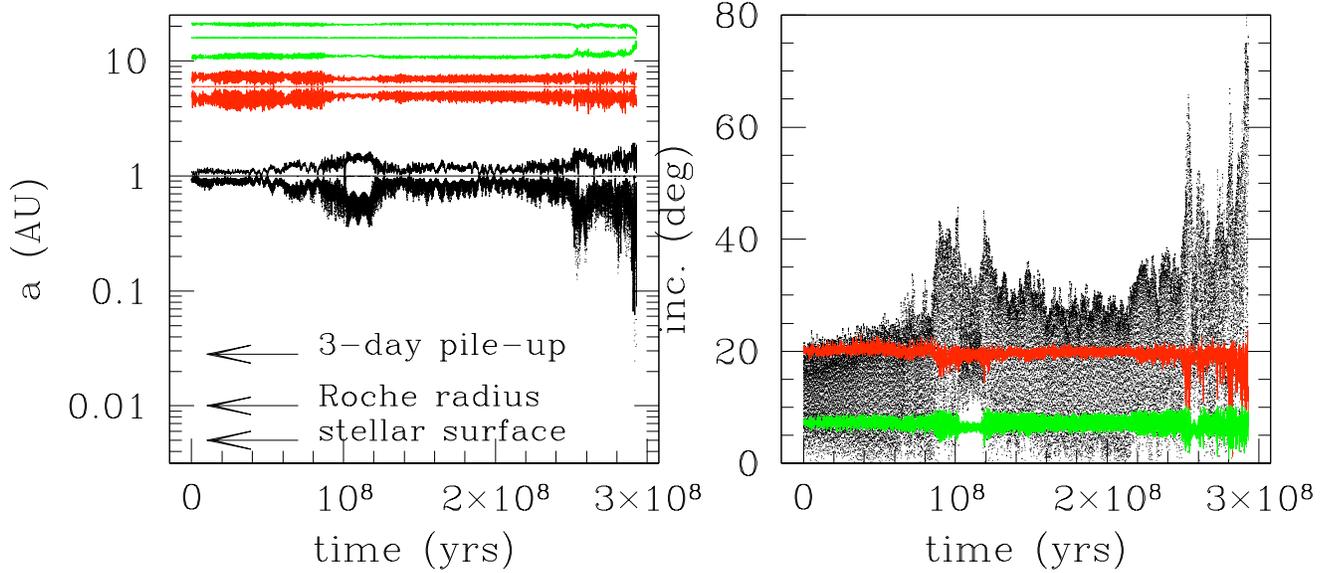}
\caption{ Formation of a hot Jupiter in our fiducial system (SWIFT
    integration with tides and GR). Left: radial excursions of the
  three planets (semi-major axis,  periapse and apoapse) are shown
  as functions of time, with the various  radii relevant for hot
  Jupiters marked  by arrows; right: planet inclinations measured
   relative to the system's invariable plane. All planets
  initially have mildly eccentric and inclined orbits, but over a
  period of $300$ Myrs, so much of the angular momentum in the
  inner-most planet can be removed that its eccentricity and
  inclination can diffusively reach order unity  values.  Planet
  interactions leave all semi-major axes unchanged, a tell-tale sign
  that secular interactions dominate the dynamics.  At $\sim 300$
  Myrs,  the pericentre of the inner planet reaches inward of a
  few stellar radii and tidal interaction with the central star kicks
  in (details in Fig. \ref{fig:largeee}).  Precessions by general
  relativity, by tidal and rotational quadruples, as well as tidal
  dissipation, prevent the pericentre  from reaching inward of the
  Roche radius. As a result, the final hot Jupiter  has  a period
  of $\sim 3$ days.}
\label{fig:textbook_case1}
\end{center}
\end{figure*}

\begin{figure*}
\begin{center}
\includegraphics[width=0.45\textwidth,angle=-90,
trim=300 100 50 110,clip=true]{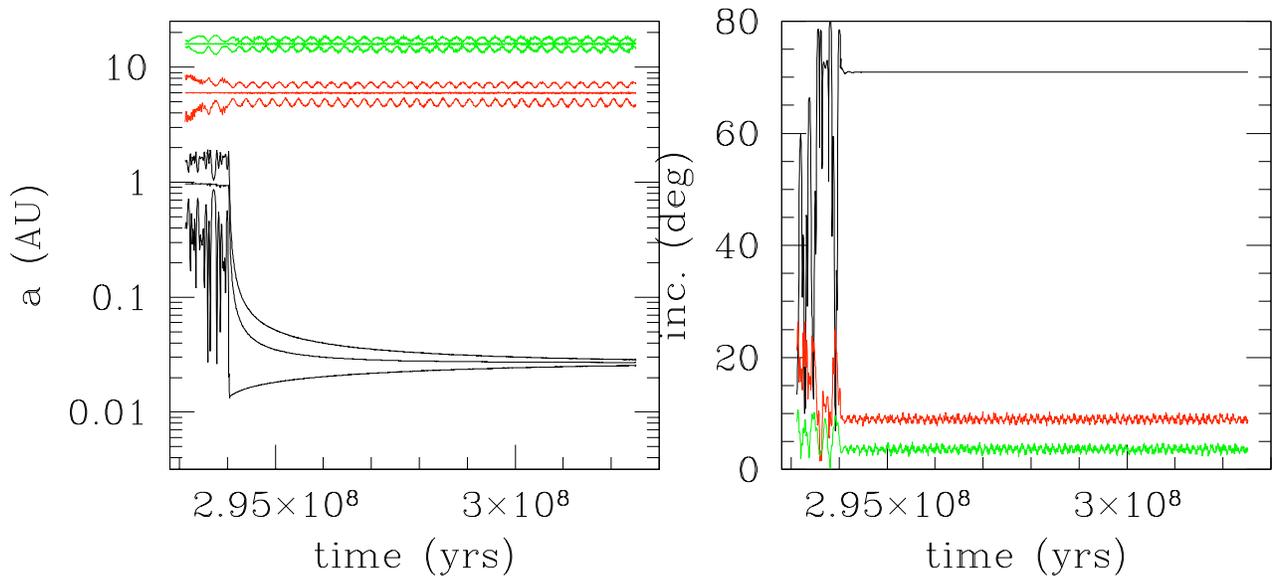}
%
\caption{Same as that in Fig. \ref{fig:textbook_case1} but expanding
  the time axis around $300$ Myrs to highlight the process of tidal
  circularization. Secular chaos  raises the inner planet's
  eccentricity diffusively to a maximum value of $0.985$,  and
    decreases its periapse to $a_1*(1-e_1) \sim 0.015$ AU -- as
  determined by a balance between secular forcing and close-range
  forces. When this occurs (at time $294$ Myrs), tidal dissipation in
  the inner planet removes orbital energy while conserving orbital
  angular momentum. This brings the planet from an initial orbit of
  $a_1 = 1$ AU to an orbit of $a_1 \approx (1-0.985^2)\times 1 {\, AU}
  \sim 0.027$ AU. It is then dynamically decoupled from the outer two
  planets. Since the total angular momentum deficit (AMD) in the
  system is absorbed by the inner planet (and subsequently removed by
  tidal dissipation), the outer two planets  lose AMD, and hence
    become more circular and coplanar after the hot Jupiter has
    formed. The inclination of the final inner orbit relative
  to the invariable plane is $\sim 70 \deg$ (right panel).  The
  evolution is highly chaotic,  and a slight modification  of
    the initial conditions changes the evolution dramatically (see
  Fig. \ref{fig:difoutcome}).  }
\label{fig:largeee}
\end{center}
\end{figure*}

{\bf Results  (Figs. \ref{fig:textbook_case1}-\ref{fig:largeee}):}
Our fiducial 3-planet system is chaotic due to secular
interactions. The random exchanges of angular momentum (but not
energy) among planets induce large fluctuations in their
eccentricities and inclinations. The orbit of the inner planet,
starting from initial values of $e_1=0.07$ and $i_1 = 4 \deg$, by
$300$ Myrs has diffused to $e_1 = 0.985$ and $i_1 = 70 \deg$.  It has
lost almost all of its  initial angular momentum to the outer
planets; equivalently, it has stolen some of the outer planets' 
angular momentum deficit.

We find that the orbital elements of the inner planet undergo a random
walk to most of the phase space allowed by the total energy and
angular momentum. However, there are a few forbidden regions. The most
important one is the region of very high eccentricity. The inner
planet {prudently} avoids the Roche zone.  This occurs because when
the pericentre of the inner planet ($a_1 (1-e_1)$) approaches the star
to within a few stellar radii, finite-size effects (quadrupole
precessions associated with the tidal and rotational deformations on
the planet and the star, respectively), as well as general relativity,
combine to suppress the secular forcing.  Were it not for these
additional precessional effects, the inner planet could be driven to
tidal disruption and merger with the central star. But as it is, the
inner planet stays within $e_1 \leq 0.985$, or $a_1 (1-e_1)\geq 0.017$
AU in our system. We explain this { in the following}.

Due to secular  interactions with other planets, the inner planet's
 longitude of pericentre precesses at the rate
\begin{equation}
{{d\pomega}\over{dt}}|_{\rm sec} \approx {{M_p}\over{M_*}} \alpha^3 n_1,
\label{eq:secforce}
\end{equation}
where $M_p$ is the mass of the perturbing planet, and $\alpha =
a_1/a_p$ is the ratio of the semi-major axes.  This is disturbed by
the prograde precession induced by the close-range forces.  Comparing
the orbit averaged rates \citep{Sterne,Shakura,WuGoldreich} for the
four types of quadrupole precessions and for GR, for the following
parameters: a Jupiter-like planet with a spin period of $3$ days,
orbiting at $a = 1$ AU around a Sun-like star that is spinning with a
period of $10$ days, we conclude that the tidal bulge raised by the
star on the planet dominates the precession at high eccentricities
with the GR effect following not far behind.  The orbit-averaged
precession due to the tidal quadrupole on the planet is
\begin{equation} {{d\pomega}\over{dt}}|_{\rm tide} = 7.5 k_2 n_1 {{(1+3
      e_1^2/2 + e_1^4/8)}\over{(1-e_1^2)^5}} {{M_*}\over{M_1}}
  \left({{R_1}\over{a_1}}\right)^5,
\label{eq:dpomegatidal}
\end{equation}
where $k_2$ is the tidal Love number of the planet, taken to be $k_2 =
0.26$, 
 and
$n_1$, $e_1$, $a_1$ and $R_1$ are the planet's mean-motion,
eccentricity, semi-major axis and radius, respectively.  Since this
rate rises steeply as the planet approaches the star, we expect that
the secular driving is arrested when the planet's pericentre reaches
inward of
\begin{eqnarray}
a_1(1-e_1) & \sim & 0.015 AU\, \left({M_1}\over{M_J}\right)^{-1/5} 
\left({M_p}\over{M_J}\right)^{-1/5} 
\left({M_*}\over{M_\odot}\right)^{2/5} \nonumber \\
& & \times \left(\alpha\over{1/6}\right)^{-3/5} \left({R_1}\over{R_J}\right).
\label{eq:bstall}
\end{eqnarray}
This stalling radius is independent of the planet's initial position
($a_1$), weakly dependent on the planetary and stellar masses, as well
as the planet spacing in the system (i.e., the value of $\alpha$). It
does, however, scale with the size of the planet linearly.  Since
Jupiter-like planets have a fairly uniform radius, the stalling
distance spans a narrow range for a wide range of system parameters.

When the orbit of such a planet is tidally circularized, the final
semi-major axis is moved to twice its stalling radius, $a_1'\simeq 2
\times 0.015 \sim 0.03$ AU.\footnote{ Angular momentum ($\propto
  \sqrt{a(1-e^2)}$) is roughly conserved during tidal dissipation. So
  the post-circularized $a_1'$ is related to the pre-circularized
  $a_1$ via $a_1'=a_1(1-e_1^2)\simeq 2 a_1(1-e_1)$.  }  This explains
why the hot Jupiters are piled up at the distance they are observed
today (Fig. \ref{fig:largeee}).

The strength of the {\it dissipative} tide can also have an affect on
the final orbit.  Even if the above precessional effects are absent,
the progression of the inner planet toward the star will be halted by
the dissipative tide, albeit at a somewhat closer distance.

\section{Analysis}
\label{sec:analy}

For secularly interacting systems, there is an important conserved
quantity, the angular momentum deficit \citep[AMD, e.g.][]{Laskar97}
\begin{equation}
  AMD \equiv \sum_{k=1}^N \Lambda_k (1 - \sqrt{1-e_k^2} \cos i_k),
\label{eq:defAMD}
\end{equation}
where $N$ is the number of planets, $\Lambda_k$ the circular angular
momentum of planet $k$, $\Lambda_k = {{m_k M_\odot}\over{m_k +
    M_\odot}} \sqrt{G (M_\odot + m_k) a_k}$, $i_k$ is its inclination
relative to the invariable plane (normal to the total angular
momentum). Since the total angular momentum is conserved, and secular
interactions do not modify the orbital energies ($a_k$ constant), the
AMD is conserved during secular interactions.  For circular, coplanar
systems, the AMD is zero, and the AMD increases with increasing $e$'s
and $i$'s.

Here, we analyze our numerical results to illustrate the condition for
hot Jupiter formation.  We find that a sufficient amount of AMD is
requisite. Firstly, the value of AMD limits the maximum eccentricity
and inclination an individual planet can attain (\S
\ref{subsec:maximum}). Secondly, only when AMD is large enough, 
  can it be shared among different secular eigenmodes
(equipartition), ultimately driving the inner planet to extreme
orbits. We call this sharing process {\bf `secular chaos'}. We
illustrate the deep analogy between AMD and kinetic energy in a
thermodynamical system in \S \ref{subsec:AMD},  and analyze the
diffusive and chaotic nature of the energy sharing process in \S
\ref{subsec:AMDdiffu} \& \ref{subsec:outcome}. We  also briefly
  look at the issue of AMD generation by mean-motion resonances in
the system (\S \ref{subsec:MMR}).

\subsection{Maximum Eccentricity and Inclination}
\label{subsec:maximum}

To be propelled into the hot Jupiter status, the inner planet has to
reach so close to the star that tidal dissipation operates.  Let this
be roughly $a_1(1-e_1) \leq 0.05$ AU. If all AMD can be transferred to
the inner planet, this condition translates into
\begin{equation}
AMD \geq \Lambda_1 
\left[1 - \left({a_1}\over{0.1 AU}\right)^{-1/2} \cos i_1\right].
\label{eq:AMDthreshold0}
\end{equation}
So a planet that is closer to the star 
  and lower in mass needs less AMD to
  become a hot Jupiter.
From now on we measure AMD in units of the circular angular
momentum of the inner planet ($\Lambda_1 = 1$). 
The above condition
  translates into
\begin{equation}
AMD \geq 1 - 0.3 \cos i_1
\label{eq:AMDthreshold}
\end{equation}
for $a_1 = 1$ AU.  So to produce a coplanar hot Jupiter ($i_1 = 0$),
$AMD > 0.7$, while a retrograde hot Jupiter would require $AMD >
1.0$. Fortunately, this is not difficult to satisfy -- the outer
planets can carry plenty of AMD even at low values of
eccentricity/inclination.  Our example system has $AMD = 1.17$.
Retrograde hot Jupiters can potentially be produced.

\subsection{AMD and Kinetic Energy}
\label{subsec:AMD}

The AMD is for a secularly interacting system what kinetic energy (or
temperature) is for a thermodynamical system. This analogy runs deep,
as we shall show here.

We introduce the complex Poincar\'e variables $z_k$ and $\zeta_k$
\citep[see, e.g.][]{Laskar97,MD00},
\begin{eqnarray}
z_k & =  & \sqrt{2} \sqrt{1-\sqrt{1-e_k^2}} \exp(i \pomega_k),  \nonumber \\
\zeta_k & = &  \sqrt{2} \sqrt{\sqrt{1-e_k^2} (1 - \cos i_k)} 
\exp(i \Omega_k),
\label{eq:poincare}
\end{eqnarray}
where $\pomega_k$ is the longitude of periapse, and $\Omega_k$ the
longitude of the ascending node.  At low eccentricities and
inclinations, $z_k \approx e_k \exp(i \pomega_k)$, and $\zeta_k
\approx \, i_k \exp(i \Omega_k)$.  The AMD may then be recast as
\citep{Laskar97}
\begin{equation}
  AMD = 
  \sum_{k=1}^N {\Lambda_k\over 2} 
  (|z_k|^2 + |\zeta_k|^2).
\label{eq:AMDinpoincare}
\end{equation}
The resemblance of AMD to kinetic energy becomes obvious in this form:
while the ``inertial mass'' for each planet corresponds to its
circular angular momentum $\Lambda_k$, the $(z_k,\zeta_k)$ pair
corresponds to its ``velocity.''

When AMD is zero, the system will remain coplanar and circular and
stable forever. At low AMD, secular interactions lead to periodic
exchanges of angular momentum between planets. This, however, is not
related to the equipartition process. The dynamics can be decomposed
into that of linear secular eigenmodes.  And the periodic variations
in orbital elements are caused by the interference between these modes
(the so-called Laplace-Lagrange theory).  Each linear mode oscillates
at its characteristic eigenfrequency with constant amplitude and
phase. Let the eigenvectors be ${\tilde z}_{k\alpha}$ and ${\tilde
  \zeta}_{k\alpha}$.  { They are orthonormal after each component is
  pre-multiplied by $\sqrt{\Lambda_k/2}$, i.e., } $\sum_k
{\Lambda_k\over 2} {\tilde z}_{k\alpha} {\tilde z}_{k\beta} = \sum_k
{\Lambda_k\over 2} {\tilde \zeta}_{k\alpha} {\tilde \zeta}_{k\beta} =
\delta_{\alpha\beta}$, where $\delta_{\alpha\beta}$ is the Kronecker
delta; { the overall phase of each eigenvector is chosen so that all
  components are real.}  Projecting the complex orbital elements onto
these eigenvectors, $z_k = \sum_{\alpha=1}^{N} a_\alpha {\tilde
  z}_{k\alpha}, \zeta_k = \sum_{\alpha=N+1}^{2N-1} a_\alpha {\tilde
  \zeta}_{k\alpha}$,\footnote{There is a trivial inclination mode with
  zero frequency, which corresponds to the overall tilt of the
  reference plane. Here, we take the reference plane to be the
  invariable plane, so the amplitude of this mode is zero.} we can
re-express AMD as
\begin{equation}
AMD  
=  \sum_{\alpha = 1}^{2 N-1} |a_\alpha|^2 .
\label{eq:AMDmode}
\end{equation}
Each eigenmode resembles one degree of freedom in a thermodynamical
system, and the AMD resembles the total energy.  In the linear
solution, if one mode is initially assigned all the { AMD}, it will
retain it forever. As a result, each planetary orbit moves within a
certain bound as given by the initial condition.


As AMD rises, energy transfer  (or really, AMD transfer) becomes
non-periodic and chaotic. Orbital elements are allowed to wander as in
a random-walk diffusion. This may ultimately lead to AMD equipartition
between different secular modes (different degrees of freedom in the
system), as well as approximate AMD
equipartition between different planets.\footnote{ { Equipartition of
    AMD amongst planets is only} approximately correct. As an example,
  the four terrestrial planets in the Solar system have similar AMD
  today. They have likely undergone extensive chaotic diffusion in the
  past.}  The least massive or the closest planet has the smallest
inertia. AMD equipartition implies that such a planet can reach very
high eccentricity and/or inclination, providing the condition for hot
Jupiter formation.

\subsection{Diffusion of AMD observed}
\label{subsec:AMDdiffu}

Diffusion of energy in a weakly nonlinear system is an extensively
studied subject, starting from the famous Fermi-Pasta-Ulam (FPU)
problem \citep{FPU}. In the following, we present  evidence for AMD
diffusion in our example system, and illustrate the criterion for AMD
diffusion.

 One line of evidence comes from  the amplitudes of the
secular eigenmodes.\footnote{Here, we decompose using the linear
  eigenvectors even though they are invalid at large amplitudes. A
  more rigorous approach may be to decompose using nonlinear
  eigenvectors, as is done in \citep{Laskar08}. But our approach
  suffices for the purpose of illustrating AMD diffusion.}
    The
initial conditions chosen in Table \ref{tab:initial} correspond to 
  deliberately depositing almost all AMD into the secular modes
associated with the outer planets, and little in the eccentricity mode
associated with the inner planet.  There are three advantages to this.
First,  AMD transfer between modes occurs on secular or longer
timescales. So for the first tens of millions of years, all three
planets have  small $e$'s and $i$'s, as appropriate for planets
  emerging from a protoplanetary disk. Second, by initializing the
inner eccentricity mode with low amplitude, we are furthest away from
our preferred end state, when this mode acquires enough  AMD to
place the inner planet on a $e\sim 1$ orbit and a tidal encounter with
the star.  Third, by concentrating  AMD into only a few modes,
we can best observe the  approach towards equipartition.

\begin{table}
\begin{center}
  \caption{Frequencies and Amplitudes (initial and final) 
    of the five  linear secular eigenmodes. }
\begin{tabular}{l|c|c|c}
\hline
mode & frequency  & \multicolumn{2}{c}{amplitude ($|a_\alpha|$)} \\
  &  (arcsec/yr) & $t=0$ & $2.9\times 10^8$ yrs \\
\hline 
e1 & 4.83 & 0.03& 0.63 \\
e2 & 8.30 & 0.31& 0.21 \\
e3 & 2.04 & 0.83 & 0.51 \\
i1 & -4.61 & 0.14 &  0.24 \\
i2 & -10.57 & 0.62 & 0.64\\ 
\hline
\end{tabular}
\label{tab:initial_mode}
\end{center}
\end{table}

\begin{figure*}[t]
\begin{center}
\includegraphics[width=0.34\textwidth,angle=-90,
trim=400 130 20 120,clip=true]{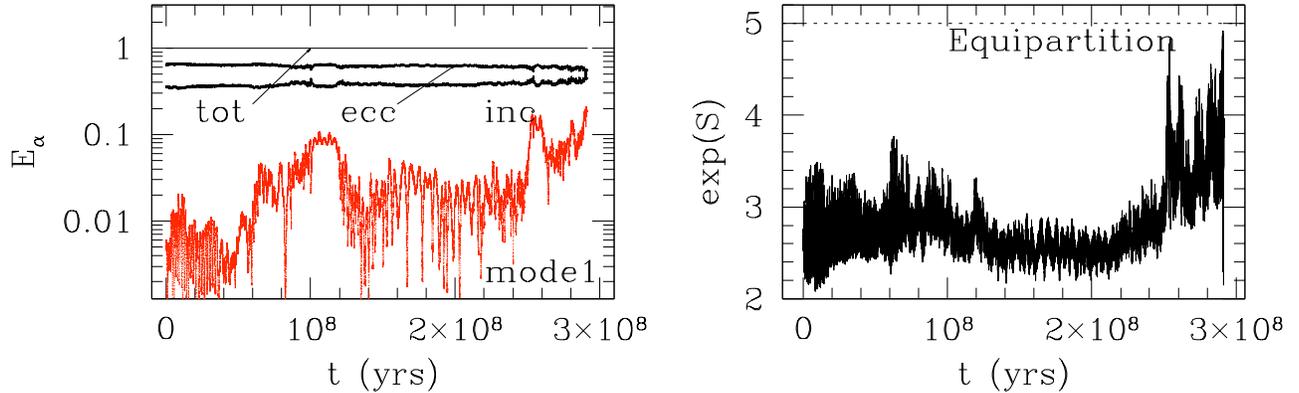}
\caption{
The diffusive behaviour for the system in
  Fig. \ref{fig:textbook_case1}, demonstrated using eigenmode energies
   (really, AMD's)
  and spectral entropy, The curves in the left panel are, from top to
  bottom, total AMD, AMD in eccentricity modes, AMD in inclination
  modes, and AMD in the eccentricity mode associated with the inner
  planet, all plotted as their ratio to the total AMD (which is
  conserved to better than a percent 
   before tidal dissipation sets in).  For clarity, the last three
  are plotted as running overages (over $2\times 10^4$ yrs).  The
  first eccentricity mode gains 
   AMD from other modes, and there is
 AMD
 exchange between eccentricity and
 inclination modes. The right hand panel shows the
exponential of the spectral entropy,
 $\exp(S)$, as
  expressed in eq. \ref{eq:entropy}.
  The process of 
   AMD
   equipartition raises $\exp(S)$ from $2$ (2
  modes dominating the 
   AMD
  ) to $\sim 5$ (all modes share comparable
   AMD,  marked by the dotted curve). 
When tidal dissipation sets in, 
   the total 
   AMD
  decreases,
   leading to a drop in the entropy.
   If we turn off the dissipation and integrate the system further, we
   observe that the spectral entropy fluctuates, but it seldom departs
   from the equipartition value for an extended period. }
\label{fig:diffusion}
\end{center}
\end{figure*}

Table \ref{tab:initial_mode} and Figure \ref{fig:diffusion} show that
after $294$ Myrs of evolution (right before tidal dissipation sets
in), the AMD of mode $1$ has grown diffusively by a factor of $\sim
20^2 \sim 400$, while the total AMD $= \sum_\alpha |a_\alpha|^2$
remains constant.  The normalization of $|a_\alpha|$ in Table
\ref{tab:initial_mode} is such that the circular angular momentum for
the inner most planet (at $1$ AU) has the numerical value of one. AMD
equipartition is reached in $\sim 2.5\times 10^8$ yrs.

A second way to quantify diffusion is by tracing the ``spectral
entropy'' \citep{Livi,Lichtenberg}
\begin{equation}
S \equiv - \sum_{\alpha=1}^{2 N-1} E_\alpha \ln E_\alpha,
\label{eq:entropy}
\end{equation}
where $E_\alpha = |a_\alpha|^2/\sum_\alpha |a_\alpha|^2$ is the
fractional energy (AMD) in mode $\alpha$.  This entropy increases from
its minimum value of $0$ (when one mode has all the AMD) to the
maximum value of $\ln N_{\rm mod} = \ln (2 N_{\rm pl}-1) = \ln 5$ as
the system diffuses toward energy equipartition.  Spectral entropy is
analogous to entropy in a thermodynamical system.

Fig. \ref{fig:diffusion} shows that for our example system, AMD is
gradually shared among different eccentricity and inclination modes,
and the spectral entropy rises toward $\ln 5$ over a $10^8$ year
timescale.

\begin{figure}[t]
\begin{center}
\includegraphics[width=0.40\textwidth,angle=-90,
trim=60 100 50 130,clip=true]{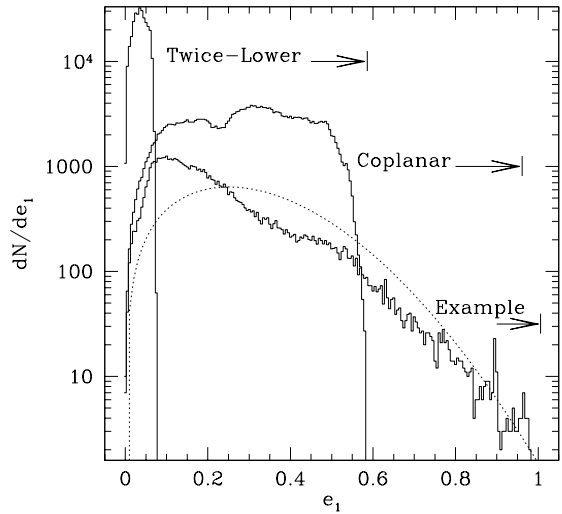}
\caption{The probability distribution of the inner planet's
  eccentricity, $dN/de_1$, in the `Example' system of
  Fig. \ref{fig:textbook_case1}, in a system with twice lower
  eccentricities and inclinations (`Twice-Lower'), and in a system
  that has the same eccentricities but is coplanar (`Coplanar'). The
  first case is integrated for $300$ Myrs (stopped short when the
  planet reaches the zone of tidal circularization), while the latter
  two are integrated for $1.5$ Gyr. The maximum eccentricity allowable
  in each case (see Fig. \ref{fig:e1_inc1}) { are marked by the
    arrows}. Diffusion in the latter two cases is partially
  inhibited. The dashed line is a Rice distribution with mean $0.2$
  and variance $0.25$. The example system is expected to approach the
  Rice distribution after a long time.}
  \label{fig:threepdf}
\end{center}
\end{figure}

A third way to observe diffusion is to plot the probability
distribution function of $e_1$, following \citet{Laskar08}, as shown
in Fig. \ref{fig:threepdf}. For our example system, $e_1$ is broadly
distributed from $0$ to $1$, with its distribution roughly described
by a Rice function \citep{Laskar08},
\begin{equation}
f(e) = {e\over{\sigma^2}} \exp\left({{-(e^2 + \nu^2)}\over{2\sigma^2}}\right)
\, I_0\left({{e\nu}\over{\sigma^2}}\right),
\label{eq:ricefunction}
\end{equation}
where $I_0$ is the modified Bessel function of the first kind with
order 0.  This is the expected distribution for $|z|$ if $z = x + i y$
with $x$ and $y$ being two independent Gaussian variables with mean
$\nu$ and variance $\sigma$.

What is  required for AMD diffusion  to occur?  We find that
 there must be a sufficient amount of AMD in the system.  For
instance, when integrating the example system, but with all secular
mode amplitudes reduced by a factor of $2$ (total AMD reduced to
$1.17/4=0.29$, the `Twice-Lower' curve in Fig. \ref{fig:threepdf}), or
when flattening all  orbits into coplanar ones (AMD$=0.72$, the
`Coplanar' curve in Fig. \ref{fig:threepdf}), we find that AMD
diffusion is largely suppressed.
While AMD allows $e_1$ to reach $0.58$ (`Twice-Lower') and $0.955$
(`Coplanar'), diffusion is only able to bring $e_1$ to $0.15$ and
$0.58$, respectively. 
Motion is either largely quasi-periodic or
weakly chaotic.
This is in contrast to the example case where the inner planet
explores phase space and reaches its maximum eccentricities and
inclinations.

Why does the amount of AMD make a difference? The interested readers
are referred to \citet{yoram} for detailed quantitative analysis. Here
we only comment that to allow AMD diffusion,  chaos is essential,
 and chaos is driven by the overlap of resonances
  \citep{Chirikov}.  The resonances of relevance are high order
secular resonances.  Their  widths increase sharply with mode
  amplitude.  A lowering of the mode amplitudes by a
mere factor of $2$  can shrink the resonance width by a large
factor. This qualitatively explains why the twice-lower case is not
chaotic. In a coplanar system, all resonances that involve the
inclination modes are ineffective,  and with fewer resonances to
  drive chaos, they dynamics become much more regular.



\subsection{Chaotic Processes Leading to Hot Jupiters}
\label{subsec:outcome}

\begin{figure*}
\begin{center}
\includegraphics[width=0.45\textwidth,angle=-90,
trim=300 100 50 110,clip=true]{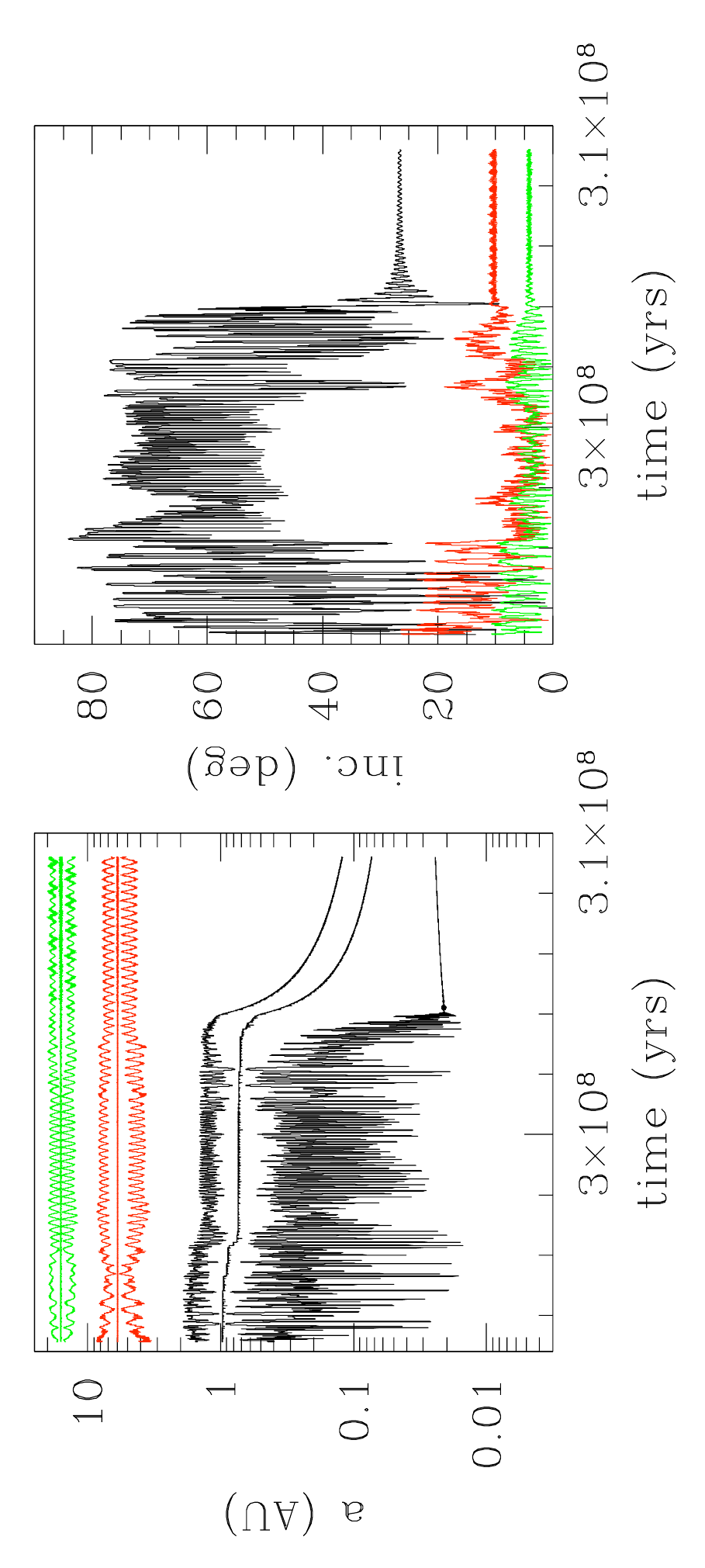}
%
\caption{Evolution of a system that is slightly modified from that in
  Fig. \ref{fig:largeee} by $0.1\%$ in total orbital energy and
  angular momentum, at time $t= 2.93\times 10^8$ yrs.  The subsequent
  trajectory differs.  The inner planet in this case 
  suffers some tidal dissipation at $t=294$ Myrs, and is migrated to
  $a_1 = 0.75$ AU.  { However, it} avoids being turned into a hot
  Jupiter right-away, and returns its AMD to the outer planets (left
  panel).  Secular chaos continues to operate until the inner planet
  is turned into a hot Jupiter with $a_1 = 0.04$ AU and an inclination
  of $25 \deg$.  This demonstrates the sensitive dependence on initial
  conditions.
}
\label{fig:difoutcome}
\end{center}
\end{figure*}

Given the chaotic nature of the dynamics, the final state of the
system depends sensitively on initial conditions. In our example case,
we obtain a hot Jupiter with $a_1 = 0.027$ AU and an inclination of
$70 \deg$ at the end of the 300 Myrs integration
(Fig. \ref{fig:largeee}). In another case, modified from the previous
one in energy and angular momentum by about $0.1\%$ at time $293$ Myrs
(just before the onset of tidal dissipation), we find that the inner
planet narrowly avoids being tidally circularized into a hot Jupiter
straight-away, and is able to return its AMD to the outer planets. The
eventual hot Jupiter thus formed has $a_1 = 0.04$ AU and an
inclination of $25 \deg$ (Fig. \ref{fig:difoutcome}). { Our other
  experiments show that in some cases the tidal dissipation process is
  gradual and occurs over many episodes of high eccentricities. There
  could be an extended period during which the planet destined to
  become a hot Jupiter is temporarily parked at an intermediate
  distance (e.g., $a_1 = 0.1$ AU) with large eccentricities and
  inclinations. They may help to explain the presence of 'warm
  Jupiters'.

With a tidal quality factor for the planet of $Q_p = 10^5$, the example
case sees most of the orbital binding energy deposited inside the
planet within a couple Myrs, with an averaged (over 1 Myrs timescale)
heating rate of $5\times 10^{29} \ergs/\s$, $\sim 15$ times higher
than the self-luminosity of a $T_{\rm eff} = 1000 \K$ Jovian planet.
The total tidal energy deposited is $\sim 1.5\times 10^{44} \ergs$,
again about $15$ times larger than the gravitational binding energy of
the planet. This has the potential of disrupting the planet unless the
heat is deposited in regions of short thermal time. However, if the
tidal dissipation process is more gradual, the mode of heat deposition
can be drastically different, potentially impacting on the final sizes
of hot Jupiters.}

\subsection{Effects of Mean-motion Resonances}
\label{subsec:MMR}

Mean-motion resonances (MMRs), not to be confused with secular
resonances, can also affect secular chaos.  This is true even if the
MMR's are of high order.  Because MMRs can change the semi-major axes,
the dynamical system has more degrees of freedom and can explore
different parts of phase space, potentially enhancing diffusion.  We
experiment by placing the outer two planets near the $2:1$ MMR (6 AU
and 9.52AU). We find that diffusion proceeds quickly even with an AMD
as low as $0.29$, a value that corresponds to our twice-lower system
(which shows regular behaviour). But when we move the outer planet
outward by a mere $0.5$ AU, the system behaves regularly again.  This
confirms that, at least in this case, MMR's can be responsible for
facilitating AMD equipartition.


In addition, even for systems which are not initially next to any
lower order MMR's (such as our example case), as eccentricity and
inclination rise to order unity values, low order MMR's are
activated. For the case shown in Fig. \ref{fig:textbook_case1}, the
semi-major axis of the inner planet undergoes increasingly large
variations at late stages of the evolution, perhaps indicating MMR's
at work. Accompanying this is the non-conservation of AMD.  Are MMR's
a significant source of AMD?  If so, they could allow the inner planet
to reach a higher eccentricity than allowed by the initial AMD of the
system.  This intriguing possibility deserves to be explored.

Another possible connection between secular chaos and MMR's exists.  In
systems which are initially more compact than our example system, but
not compact enough to have immediate close encounters, secular chaos
increases the eccentricity of inner planets, allowing MMR's to function
at later stages, leading finally to planet scattering well after the
protoplanetary disks have dissipated. 



\section{Predictions for Hot Jupiters}
\label{sec:observation}

We have demonstrated that secular chaos can produce hot Jupiters. The
remaining central issue is how prevalent this mechanism is.  { Is it
  prevalent enough to explain the observed frequency of hot Jupiters?}

Unfortunately, since we do not know the initial configurations of
planetary systems, it is difficult to predict the frequency of hot
Jupiter production by secular chaos. We also face the problem that a
systematic survey of the relevant parameter spaces is numerically
expensive, at least using our current technique of N-body integration.
So in the following, we discuss qualitative predictions based on our
present understanding of secular chaos.

\subsection{General Predictions}

The predictions are ranked roughly in  decreasing order of certainty.

\begin{enumerate}

\item a pile-up of hot Jupiters around three-day orbital periods. 

  The characteristic stalling at $3$-day orbital periods that we
  observe in our simulations is explained by a combination of tidal
  precession and tidal dissipation (both due to tides on the
  planet). Precessions by other close-range forces (GR, rotational
  quadrupole) are less important.  The tidal precession stalls the
  rise of the eccentricity to about a few times the Roche radii, and
  tidal dissipation finishes the job by circularizing the highly
  eccentric orbit to that of a hot Jupiter.  Location of the pile-up
  remains largely unchanged even if the rate of tidal dissipation is
  orders of magnitude weaker.  This conclusion also applies to Kozai
  migration, another secular migration process.

\item hot Jupiters are lower in mass compared to other giant planets.

  This prediction stems from AMD conservation. To become a hot
  Jupiter, the inner planet has to reach an eccentricity so high that
  $a_1(1-e_1) \leq 0.05$ AU. For a given amount of AMD, a lower mass
  inner planet can reach a higher eccentricity.  Compared to giant
  planets at larger distances, there is a clear deficit of massive hot
  Jupiters \citep{Zucker,Udry}. Kozai migration, in contrast, has no
  mass preferences \citep{WuMike}.

\item hot Jupiters have no companion within a few AU, but have
  companions roaming at larger distances.

  On the one hand, the high eccentricity episode the inner planet
  undergoes before it is tidally captured implies that hot Jupiters
  have no companions within a few AU. This agrees with observations
  where hot Jupiters appear to be strikingly alone \citep{Wright}, and
  where attempts at measuring transit timing variations have
  repeatedly turned up empty-handed. On the other hand, secular chaos
  requires driving by other giant planets. They should be roaming at
  large distances (outside a few AU) and remain to be detected -- some
  may have already shown up as radial velocity  residuals in hot
  Jupiter systems \citep{Fischer,Wright}. There is also the intriguing
  possibility that one of the companions is a binary star.

\item frequency of hot Jupiters should rise with stellar age.

  Being a diffusive process, secular chaos operates on timescales
  comparable  to or longer than the secular precession timescale. The
  latter, for our fiducial system, is of order $M_*/M_2 P_2^2/P_1 \sim
  10^5 $ yrs. As such, we expect stars that are within a few tens of
  million years after their disk dispersal should have a lower hot
  Jupiter fraction than stars that are a few Gyrs old. This prediction
  should be quantified by extensive numerical simulations that start
  with reasonably realistic planetary configurations.

\item orbits of hot Jupiters could be strongly misaligned with stellar
  spin.

We will discuss the planet inclination in \S \ref{subsec:obliquity}. 

\end{enumerate}

If secular chaos is responsible for producing hot Jupiters, this has a
number of implications for planetary systems at large.

\begin{enumerate}

\item a large number of planetary systems should have 3 or more giant
  planets on (mildly) eccentric, inclined orbits.

  Only systems that have sufficient AMD
  can   make hot Jupiters. Hot Jupiters may be the
  tip of the iceberg in terms of their system AMD values. If so, then
  most giant planets we know of should reside in systems with three or
  more giant planets. There should still be residual AMD in the outer
  planets.

\item warm Jupiters could arise from secular chaos and their orbit
  could be misaligned with respect to the stellar spin.

  We call giant planets at a few  tenths of an AU `warm Jupiters,'
  for the reason that they are at intermediate locations between hot
  Jupiters ($\sim 10^{-2}$ AU) and cold Jupiters ($\sim$ a couple AU).

  Among those that fail to make hot Jupiters, there may be warm
  Jupiters -- planets which have suffered some degree of tidal
  dissipation but have yet to be tidally captured into hot Jupiters.
  These planets are temporarily intercepted from their inward spiral
  by interactions with outer planets that can inject into their orbits
  a fresh boost of angular momentum.  Eventually, these planets would
  be dragged in to become hot Jupiters, but while they are in their
  temporary parking space, their orbits could have high inclinations
  as well as high eccentricities. Observationally, there is a `period
  valley' at these distances, indicating that perhaps the warm Jupiter
  phase is relatively short-lived. Measurements of spin-orbit angle
  for these planets will be useful constraints.


\item even for systems where the inner planets can not reach high
  enough eccentricity to be captured into  hot Jupiters, secular chaos
  may have observable consequences.
  
  Starting from mildly eccentric, inclined orbits, inner planets in
  planetary systems may gradually extract AMD from outer planets. The
  long-term eccentricity and inclination distributions for these
  planets approach the Rice distribution (Fig. \ref{fig:threepdf}).
  This reduces to a Rayleigh distribution when the centroid is much
  smaller than the variance. Fig. \ref{fig:edist} shows the observed
  eccentricity distribution of Jovian planets: it may be represented
  by a Rayleigh distribution with variance $0.25$. If future
  Rossiter-McLaughlin measurements provide us with a similar
  distribution for the orbital inclinations, this will be a strong
  proof in favour of secular chaos.

\begin{figure}[t]
\begin{center}
\includegraphics[width=.5\textwidth]{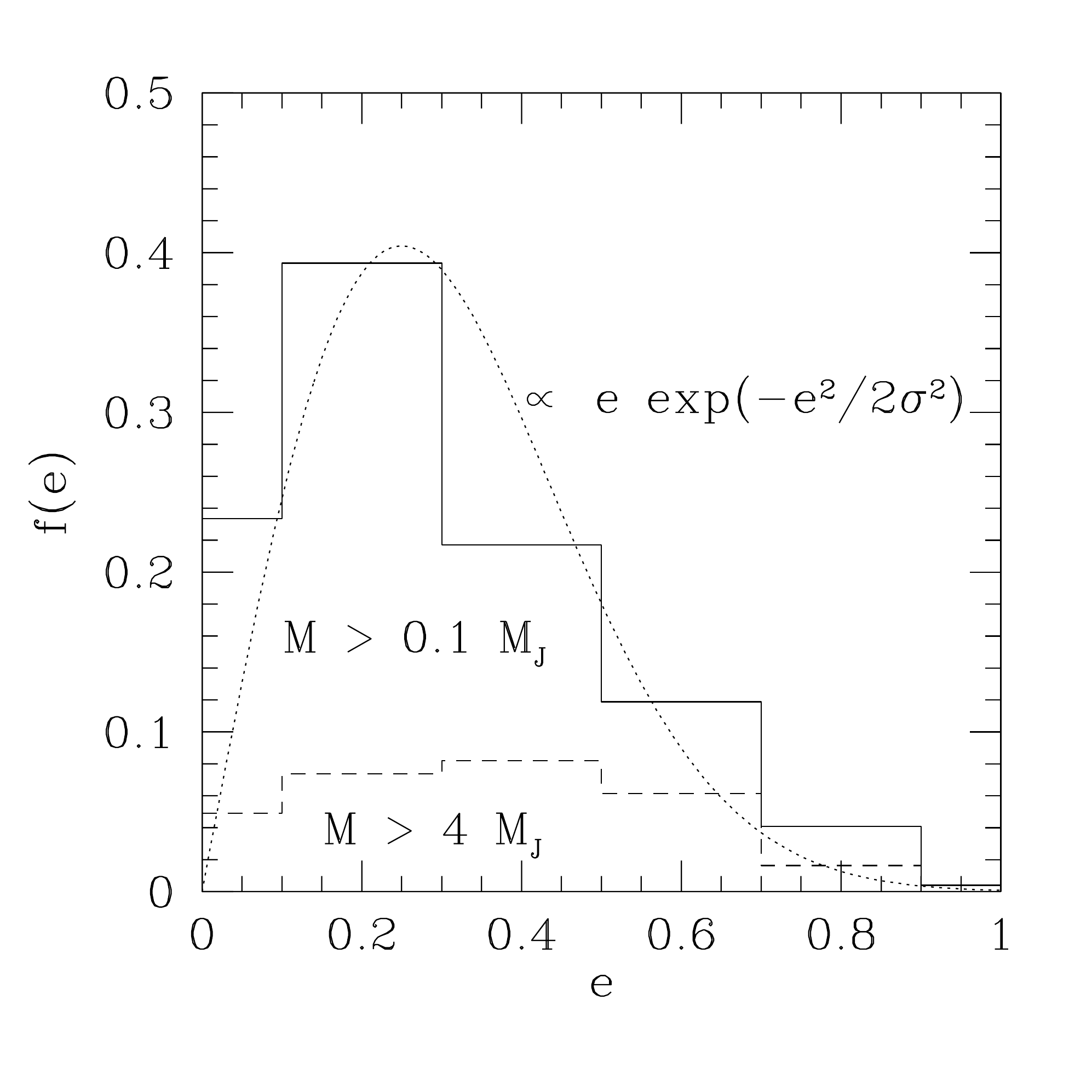}
\caption{The eccentricity distribution of Jovian planets that have
  pericentre between 0.1 and 10 AU and projected masses above $0.1
  M_J$. (the solid  histogram). The lower limit of 0.1 AU is selected to
  preclude planets that have undergone significant tidal
  circularization. The dotted line is the Rayleigh distribution with
  $\sigma = 0.25$. The sub-population of more massive planets with $M
  \sin i > 4 M_J$ are represented by the dashed histogram. It appears
  that more massive planets have a hotter distribution. }
\label{fig:edist}
\end{center}
\end{figure}

If secular chaos is largely responsible for the observed planet
eccentricities, one expects that, for the same amount of AMD available
in the system, lighter planets in general reach higher
eccentricities. However, data (Fig. \ref{fig:edist}) portray an
inverse correlation -- more massive ($M\sin i > 4 M_J$) planets have a
hotter eccentricity distribution.  This may be interpreted as due to
pollution from systems undergoing Kozai cycles induced by stellar
companions, which is capable of exciting eccentricities even in very
massive planets. This interpretation requires that massive planets
occur preferentially in binary systems.



\item secular chaos can stabilize planetary systems.

  In our simulation (Fig. \ref{fig:textbook_case1}), the tidally
  captured hot Jupiter removes AMD from the system, stabilizing the
  outer planetary systems as a result. This could be a generic process
  in organizing planetary systems on long timescales.

\item can secular chaos explain hot Neptunes or hot Earths?

  If the inner planet has a lower mass, it experiences a weaker
  near-range precession due to its smaller tidal and J2 moments.
  Eq. \ref{eq:bstall} predicts that a Neptune-like planet ($M = M_N =
  1/17 M_J$, and $R = R_N = 0.36 R_J$), driven to chaos by secular
  forcing of giant planets at a few AU, could be stalled at
    periapse distance $\sim 0.009 AU$, and hence it would be
    circularized at $\sim 0.018 AU \sim 4 R_\odot$.  This distance
    would be approximately doubled if it is forced by other
    Neptune-mass planets, while a more inflated planet would also be
    stalled at a greater distance.  
  Hot Neptunes thus produced will tend to have a broader pile-up than
  hot Jupiters, because the masses of the perturbers can extend  over a
    large range.  Such a mechanism could explain some of the observed
  hot Neptunes -- for instance, the HD 125612 system where a hot
  Neptune is accompanied by two Jupiter-like planets at much larger
  distances \citep{LoCurto}.

  \citet{Bouchy} noted that $\sim 70\%$ of low mass close-in planets
  (with $M \sin i < 0.1 M_J$) have detected planetary companions. This
  differs from the hot Jupiter case.  It likely relates to the lower
  requirement for the planetary spacing when the inner planet is less
  massive.


  An Earth-like planet, on the other hand, will not be stalled at
  large enough distances to avoid Roche-lobe overflow. Secular chaos
  is an unlikely agent for making hot earths, especially considering
  that the tidal damping time for an Earth-like planet at $0.03$ AU
  exceeds a Hubble time. 
  


\end{enumerate}

\subsection{Stellar Obliquity}
\label{subsec:obliquity}

\begin{figure}[t]
\begin{center}
\includegraphics[width=0.48\textwidth,angle=-90,
trim=50 100 30 100,clip=true]{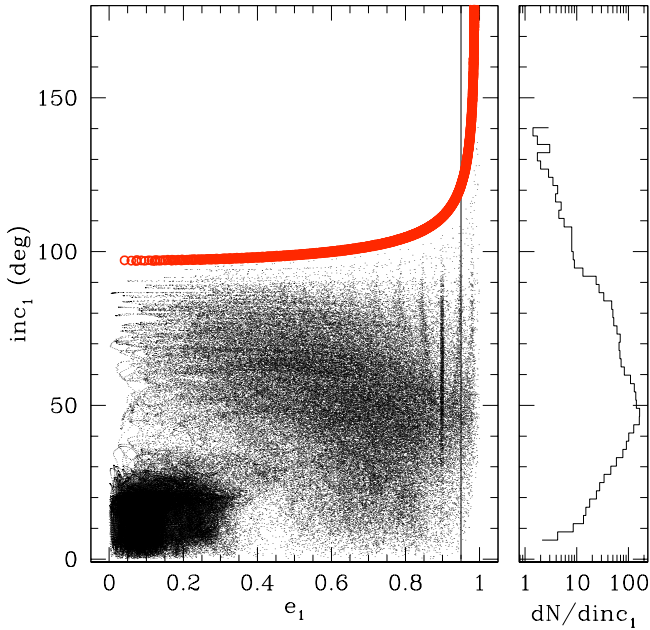}
\caption{The eccentricity-inclination phase space that is traversed by
  the same system as Fig. \ref{fig:textbook_case1} over a total of
  $400$ Myr, in the absence of tidal dissipation.  The thick red curve
  bounds the region within which the inner planet can explore, as
  dictated by energy and angular momentum conservation (assuming
  constant $a_i$, so this is determined by the initial AMD).  Secular
  chaos transports the inner planet to very high eccentricities
  (bounded at the right by near-range precessions), and to retrograde
  orbits.  If tidal dissipation were turned on, it would operate
  rightward of the vertical line at $e=0.95$ (pericentre distance of
  $a_1 (1-e_1) = 0.05$ AU).  The right panel shows the distribution of
  inclinations whenever $e_1 > 0.95$.  This can be regarded as an
  approximation to the distribution of final inclinations.  For this
  system, retrograde orbits are achieved a few percent of the time.
}
\label{fig:e1_inc1}
\end{center}
\end{figure}

The final inclination of the inner planet deserves an in-depth
discussion.

\citet{FabryckyWinn}  found that a majority of transiting planets
have their orbits aligned with the stellar spin.   However, that
  majority was quickly weakened by a slew of misaligned systems
\citep{Triaud}. The current situation is unclear, partly due to small
number statistics,
though a correlation between stellar obliquity and stellar spectral
type has been noted \citep{Winnetal}.


Our example system (Fig. \ref{fig:textbook_case1}) produces a hot
Jupiter with an orbit inclined by $70 \deg$ from the invariable plane
of the planetary system, while its near-identical twin yields a hot
Jupiter with an inclination of $25 \deg$ (Fig. \ref{fig:difoutcome}).
Secular chaos could in principle drive the inner planet to
inclinations between $0$ and $180 \deg$ relative to the invariable
plane (Fig. \ref{fig:e1_inc1}), given the amount of AMD in this
system.
To demonstrate this, we turn-off tidal dissipation in our example
integration, and evolve the fiducial system for $400$ Myr. The
eccentricity and inclination phase-space traversed by the inner planet
is presented in Fig. \ref{fig:e1_inc1}. Within the constraint of AMD
conservation, the planet is able to diffusively reach high
eccentricity,  with both prograde and retrograde orbits. The final
inclination for the hot Jupiter then depends on chance.

In Fig. \ref{fig:e1_inc1}, we also display the distribution of
inclinations whenever $1 - e_1 \leq 0.05$, as a proxy for the final
inclinations of hot Jupiters. We find that retrograde orbits represent
a few percent of the total.  The preponderance of prograde orbits is
related to the prograde initial condition.

This result is specific to our example system and is relative to the
invariable plane.  The invariable plane does not necessarily coincide
with the stellar equatorial plane -- the relative angle is only $7
\deg$ in the Solar system, but it can be large
elsewhere.\footnote{Here, we have assumed that the stellar spin is
  aligned with the invariable normal when calculating precession due
  to the stellar rotational bulge.}

\section{Summary}
\label{sec:conclusion}

Hot Jupiters, while representing only a small fraction of all known
extra-solar planets, demand special attention. They are most at odds
with planet formation theory; they are detected disproportionately in
radial velocity and transit surveys;  and they are most accessible
to characterization. Their rarity may indicate that their formation
requires extreme circumstances.   However, they may teach us much
about the general conditions of planetary systems.

Hot Jupiters are piled up around 3-day orbital periods, with rapid
cut-offs both inward and outward of this distance;  they tend to
  be less massive than more distant planets; many of them have orbits
that are misaligned relative to the stellar spin;  and they are
remarkably anti-social in having few detected companions.



In this work, we show that most of these  characteristics can be
explained if hot Jupiters are produced by secular chaos.  These
planets, originally located at  $\gtrsim 1$ AU, acquire AMD from
planets that are further out in the system. The outer planets can be
mildly eccentric and/or inclined. But the same AMD  produces much
  greater eccentricities and inclinations when it is transported to an
  inner planet, especially if the inner planet is less massive than
  the outer ones.  The extreme high eccentricity allows  the inner
  planet to reach inward of a few stellar radii and be tidally
ensnared by the central star into a hot Jupiter.\footnote{Kozai
  migration is another process that may give rise to large
  eccentricity to Jovian planets.  Such planets may be similarly
  captured into hot Jupiters.  Since Kozai oscillation is also a
  secular forcing, we propose to name both secular chaos and Kozai
  migration generically as ``Secular Migration''.}  We find that the
criterion for hot Jupiter production is a sufficient amount of AMD.

Only 5 hot Jupiters have known planetary companions, and these
companions are typically eccentric and may have contributed to secular
chaos. We note that the hot Jupiter (at $0.06$ AU) in the Ups And
system, where two {other} massive planets orbit at $0.8$ and $2.5$ AU,
with eccentricities $0.2$ and $0.3$ respectively, may well be produced
by the secular chaos presented here. This possibility is further
boosted by the recent finding that the outer two planets'
{inclinations} are misaligned by $\sim 30 \deg$ from each other
\citep{McArthur}.  Similarly, the retrograde hot Jupiter in WASP-8b
\citep{Queloz} is in a stellar system including an M-star companion at
$\sim 600$ AU. Moreover, the radial velocity trend indicates a
companion at distance $> 1 AU$ that is more massive than $2 M_J$. This
could also be a hot Jupiter produced by secular chaos, with the M-star
acting as the third planet.

Hot Neptunes may also be formed via secular migration. However, hot
Earths  are different . It is not that Earth-like planets could
not undergo secular chaos, but that once they do, they  cannot be
stalled at a safe  enough distance  to avoid being swallowed
by the central  star.

Secular chaos has also been found to be responsible for instability in the
inner Solar System \citep{Laskar08,yoram}.  We speculate that secular
chaos may be a frequent phenomenon in planetary systems. It may help
to excite inner planets to higher eccentricities or inclinations. If
these planets are removed, the remaining planetary systems may be
stabilized for a time comparable to the system age.

The success of this theory depends on two unknown factors.  One is the
amount of initial AMD in the system. The other is the typical
configuration of planetary systems when emerging out of the
protoplanetary disk. It also needs to be demonstrated solidly that
secular chaos can lead to a large fraction of retrograde hot Jupiters.
Observationally, if not only hot planets, but also warm or cold
planets can be shown to have significant orbital inclinations relative
to the spin of their host stars, this would boost the case for the
ubiquity of secular chaos.  Future Rossiter-McLaughlin measurements
should be extended to transiting planets at large distances.

\begin{acknowledgements}
  We thank Peter Goldreich, Jeremy Goodman, Norm Murray, Daniel
  Fabrycky, Scott Tremaine, Eiichiro Kokubo and Doug Lin for
  discussions. We both enjoyed the hospitality of KIAA Beijing where
  part of this work is performed.  YW further acknowledges the NSERC
  funding and a Caltech sabbatical stay. Lastly, we thank M. Duncan
  and H. Levison for the use of their SWIFT package.
\end{acknowledgements}

\newpage

\begin{appendix}

  The standard SWIFT package distribution does not contain a treatment
  for the general relativistic precession, or precession due to tidal
  and rotational bulges. These dominate over secular precession by
  other planets when the inner planet reaches very close to the host
  star. Together with tidal dissipation, 
   these effects determine the final
  orbit of the planet, as well as the timescale for tidal
  circularization. Below is our implementation of all these processes
  in the SWIFT code.  Since non-Keplerian accelerations in the
  symplectic SWIFT integrator are incorporated as velocity kicks
  between Keplerian drifts, we need expressions for the perturbative
  accelerations.

  To first order in $(v/c)^2$, the GR effect can be written as a
  perturbation to the Newtonian gravitational potential as
\begin{equation} 
\Phi_{\rm GR} = - {{G M_* L^2}\over{ c^2 m_p^2      r^3}},
\label{eq:phiGR}
\end{equation}
where $r$ is the radial distance between the star and the planet, $L =
m_p \sqrt{G M_* a (1-e^2)}$ is the orbital angular momentum.  The
purely radial force associated with this potential,
\begin{equation}
 {\bf a}_{\rm GR} = - {
  \bnabla} \Phi_{\rm GR} = - {{3 G^2 M_*^2 a (1-e^2)}\over{c^2 r^4}}
\, {\bf\hat r},
\label{eq:agr}
\end{equation} 
gives rise to a precession of the eccentricity vector.  We confirm
that such a numerical procedure yields the following orbit-averaged
precession rate for the longitude of pericentre  \citep{Einstein}
\begin{equation} {\dot
  \pomega_{\rm GR}} = {{3 G M_* n}\over{c^2 a (1-e^2)}}.
\label{eq:agr_average}
\end{equation}

 For the tidal effects, we first consider the rotational bulge and
the tidal bulge on the star. The following expressions are from
\citet{Sterne}, differing only in notation. Let the stellar spin rate
be $\omega_*$, radius $R_*$.   The centrifugal potential due
  to the stellar spin, and the quadrupole tidal potential  due to
  the planet acting on a point at distance ${\bf D}$ away from the
stellar center is
\begin{equation}
\Phi_{\rm acting} = + {1\over 3} \omega_*^2 D^2 P_2(\cos\theta') -
{{G m_p}\over{r}}  \left({D\over r}\right)^2 P_2 (\cos\theta).
\label{eq:2potential}
\end{equation}
Here, the Legendre function $P_2 (x) = 1/2\, (3 x^2 -1)$, and the
connecting angles are defined as $\cos \theta = {\hat {\bf D}} \cdot
{\hat {\bf r}}$, $\cos\theta' = {\hat{\bf D}} \cdot {\hat {\bf
    \omega_*}}$, where ${\bf r}$ is the vector connecting the two
bodies.  The global distortion of the star under the
  above potential casts a response potential in its surrounding
(measured at position ${\bf D}$)
\begin{equation}
  \Phi_{\rm reponse} =  {{k_{2*} R_*^5}\over{D^3}} 
  \left[   {{\omega_*^2}\over 3} P_2(\cos\theta') 
-{{G m_p}\over{r^3}} P_2(\cos\theta)  \right],
\label{eq:responsepotential}
\end{equation}
where $k_2$ is the Love number and is taken to be $0.029$ for the star
(polytrope $n=3$) and $0.52$ for the planet (polytrope
$n=1$).\footnote{It is larger than the apsidal motion constant in
  binary studies (also written as $k_2$) by a factor of $2$.} The
barycentric acceleration the reduced particle feels is therefore
\begin{equation}
{\bf a} = - {{m_p}\over{\mu}} \bnabla \Phi_{\rm response},
\label{eq:reducedmass}
\end{equation}
where the reduced mass $\mu = m_* m_p/(m_* + m_p)$. 

We trace the planetary motion in the frame of the invariable plane.
And we assume that the stellar spin coincides with the normal of the
invariable plane. So at the heliocentric position of the planet
($x,y,z$), $\cos\theta = 1$, and $\cos\theta' = z^2/r^2$, where $r^2 =
x^2 + y^2 + z^2$. Here the $z^2$ term in the potential gives rise to
spin-orbit coupling and the orbit normal of the planet precesses
around the spin direction.

The bulges on the planet are treated similarly, except we assume that
the planet spin is aligned with the orbit normal, so $\cos\theta' =
0$. 

The above close-range accelerations are expressed for the barycentric
movement.   However, one should correct for the fact that
the kick asked by the SWIFT code is the heliocentric value.

The dissipative part of the tidal effect is handled in a way that
differs from the standard treatment of \citet{LeePeale}. We use the
weak friction prescription for the equilibrium tidal bulge, and
calculate the effect of dissipation in both the star and the
planet. The tidal bulge raised on either body produces a potential {
  given by the second term in} eq. \ref{eq:responsepotential}. Due to
finite dissipation inside the body, there is a delay between the
response and the forcing. We can assume either a constant time lag
($\tau$), or a constant phase lag ($\epsilon$). In the latter case, we
can introduce a tidal $Q$ factor \citep{GoldreichSoter}, which is
related to the phase lag and the time lag as $\epsilon = 1/Q$, and
$\tau= {1\over Q}\, {{2 \pi}\over{\omega_{\rm tide}}}$,
respectively. Here $\omega_{\rm tide}$ is the tidal forcing
frequency. For the eccentricity tide, $\omega_{\rm tide}$ is simply $2
n$.  The acceleration associated with the delayed tidal bulge raised
on body $M$ (with radius $R$) by body $m$ at distance $r$ is
\citep{Hut}
\begin{equation}
 - {{Gm^2}\over{\mu r^2}}
  \left({R\over r}\right)^5 k_2 \left[ \left(3 + 9 {{\dot r}\over r}
      \tau\right) {\bf\hat r} - (\omega - {\dot \theta}) \tau 
{\bf\hat \theta}\right],
\label{eq:ffriction}
\end{equation} 
where $\omega$ is the rotational velocity and ${\dot \theta}$ the
instantaneous orbital angular velocity. In particular, ${\dot r} = n a
e \sin f/\sqrt{1-e^2} $ for a Keplerian ellipse.  The first half of
the radial term contributes to orbital precession (dealt with above)
but no energy dissipation (as it is anti-symmetric within a Keplerian
ellipse and cancels out over an orbit). We ignore the angular force
which transfers angular momentum -- the spin angular momentum of the
planet is much smaller { than its orbital angular momentum} and so we
assume that the planet is quickly synchronized with the orbit, while
we assume that the tidal $Q$ factor associated with the star is so
large that there is no angular momentum being transported between the
orbit and the stellar spin. So we are left only with the second half
of the radial force. This is easily implemented in the SWIFT
package. We adopt values of $Q_p = 10^5$ and $Q_* = 10^{10}$ for our
work. So tidal dissipation is dominated by that inside the planet.

\end{appendix}

\end{document}